\documentclass{article}

\usepackage{PRIMEarxiv}

\usepackage[utf8]{inputenc} 
\usepackage[T1]{fontenc}    
\usepackage{hyperref}       
\usepackage{url}            
\usepackage{booktabs}       
\usepackage{amsfonts}       
\usepackage{nicefrac}       
\usepackage{microtype}      
\usepackage{lipsum}
\usepackage{fancyhdr}       
\usepackage{graphicx}       
\usepackage{adjustbox} 
\usepackage{subcaption}
\title{Fine-Tuning Language Models for Context-Specific SQL Query Generation

}

\author{
  Amine Rebei \\
  DinMo \\
  Paris, France \\
  \texttt{amine@dinmo.com} \\
}

\begin{document}
\maketitle

\begin{abstract}
The ability to generate SQL queries from natural language has significant implications for making data accessible to non-specialists. This paper presents a novel approach to fine-tuning open-source large language models (LLMs) for the task of transforming natural language into SQL queries within the retail domain. We introduce models specialized in generating SQL queries, trained on synthetic datasets tailored to the Snowflake SQL and GoogleSQL dialects. Our methodology involves generating a context-specific dataset using GPT-4, then fine-tuning three open-source LLMs—Starcoder Plus, Code-Llama, and Mistral—employing the LoRa technique to optimize for resource constraints. The fine-tuned models demonstrate superior performance in zero-shot settings compared to the baseline GPT-4, with Code-Llama achieving the highest accuracy rates, at 81.58\% for Snowflake SQL and 82.66\% for GoogleSQL. These results underscore the effectiveness of fine-tuning LLMs on domain-specific tasks and suggest a promising direction for enhancing the accessibility of relational databases through natural language interfaces.
\end{abstract}


\section{Introduction}

The transformation of natural language into SQL queries has been of great interest to both academia and industry due to its potential to enable non-specialists to access data from relational databases using natural language\cite{cai2017}\cite{Qin2022ASO}\cite{Xu2017SQLNetGS}\cite{Yaghmazadeh2017SQLizerQS}\cite{Yu2018TypeSQLKT}. Recent advances in neural models, such as those based on large language models (LLMs), have led to impressive results on benchmarks like Spider\cite{Yu2018SpiderAL} and WikiSQL\cite{Zhong2017Seq2SQLGS}. For instance, the accuracy of the best-performing model on the Spider leaderboard has increased from 53.5\%\cite{Zhong2020GroundedAF} to 86.6\%\cite{Gao2023TexttoSQLEB} in the last three years. The most recent state of the art (SOTA) parsers in Spider leverage the powerful comprehension and coding capabilities of an LLM.\\ Until recently, sequence-to-sequence approaches that use language models have been the state of the art. Medium-sized models, such as T5\cite{Raffel2019ExploringTL}, have achieved good results by incorporating SQL-specific designs and domain knowledge through fine-tuning. Notable models, such as PICARD\cite{Scholak2021PICARDPI}, have employed methods like incremental parsing, relation-aware self-attention and skeleton-aware decoding frameworks.\\ Large language models, like GPT-4\cite{OpenAI2023GPT4TR}, and PaLM-2\cite{Anil2023PaLM2T}, have also shown remarkable success with zero-shot and few-shot prompting, or in-context learning\cite{Wei2022EmergentAO}. The advantage of few-shot prompting is that it requires no training, has lower computation requirements, is less likely to over-fit to train data, and is easy to adapt to new data. However, the performance may be sub-optimal compared to fine-tuned alternatives. For example, CodeX\cite{Chen2021EvaluatingLL} and ChatGPT\cite{Liu2023ACE} have demonstrated successful results with in-context learning for Text-to-SQL, but still have a clear gap with fine-tuned models that can also benefit from few-shot prompting.\cite{Sun2023SQLPaLMIL}\\ There is a great potential in open-source language models, with recent advances in programming, mathematical reasoning, and text generation tasks. However, previous research on Text-to-SQL has mainly focused on OpenAI models, overlooking the capabilities of open-source models. While open-source models may lack the functionality of their commercial counterparts in understanding context and generating coherent responses, they can be fine-tuned to improve their Text-to-SQL performance through supervised learning.\cite{Gao2023TexttoSQLEB}\\ The first attempts to finetune open-source LLMs for Text-to-SQL have yielded promising results \cite{defog}\cite{numbersstation2023NSText2SQL}, surpassing the OpenAI models on this task with a fraction of the size.\\
This paper introduces a technique for constructing Text-to-SQL models tailored to a particular context. We specialize the models to a retail context with Snowflake SQL and GoogleSQL, and demonstrate that they outperform GPT-4 in zero-shot tests. This demonstrates the ability of these models to adjust to different SQL dialects and to modify the syntax of the queries.\\
We explain the process of generating a synthetic dataset of Text-to-SQL data in a retail setting, as well as the base pre-trained models used and the techniques used to finetune them. The results are then presented and discussed.

\section{Methods}
The input for Text-to-SQL is a prompt that includes a natural language query, the details of the database, and the corresponding SQL dialect. The output is the SQL query that answers the question. The database information typically includes multiple tables, each of which has a name, columns, and data types for the columns. 
\subsection{Dataset description}
The initial segment of the dataset is comprised of a compilation of publicly accessible datasets, namely Spider \cite{Yu2018SpiderAL} and Bird-SQL \cite{Li2023CanLA}. The Spider dataset is a comprehensive and diverse semantic parsing and text-to-SQL collection, annotated by 11 Yale students, which aims to facilitate large-scale, complex, and cross-domain research. The BIRD (BIg Bench for LaRge-scale Database Grounded Text-to-SQL Evaluation) dataset is specifically developed to address the gap between academic studies and practical applications in text-to-SQL parsing. It highlights emerging challenges, such as incorporating external knowledge, handling inconsistent data, and optimizing SQL efficiency.\\
Utilizing SQLGlot \cite{sqlglot}, these datasets were transformed into Google SQL and Snowflake SQL formats. Both versions, Google SQL and Snowflake SQL, were incorporated into the training process, enabling the model to discern the distinct syntax of each.
Furthermore, we employ the GPT-4 model(the 0314 version)\cite{OpenAI2023GPT4TR} to create a specialized Text-to-SQL retail dataset. \\
\subsection{Dataset generation}
\begin{figure}[h]
  \centering
  \includegraphics[width=\linewidth]{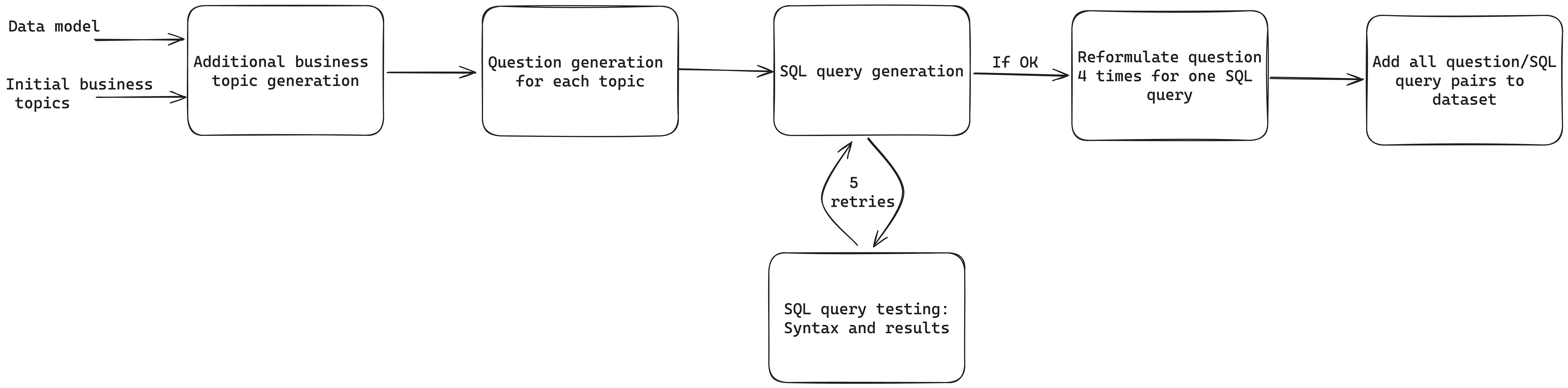}
  \caption{Text-to-SQL Data generation flow}
  \label{flow}
\end{figure}

Initially, we establish two representative retail datasets encompassing tables for customers, products, orders, and order lines. One dataset includes additional tables for sellers and payments. After their creation, these datasets are uploaded to the Snowflake and BigQuery data warehouses.\\
Subsequently, we delineate five business themes to serve as the foundation for question generation and instruct the GPT-4 model to generate additional themes. We also generate supplementary topics for the dataset that contains payment and seller information. This process results in 90 shared topics between datasets, and an extra 10 topics. These themes span a diverse range from customer demographics to seller performance, order seasonality, and product profitability, among others.\\
Following that, we prompt the GPT-4 model to generate up to 10 questions per topic that can be addressed using the provided data model, with instructions to cease generation if the questions become overly repetitive. Given that the information available for each dataset varies, this process is conducted independently for each dataset.\\
From the preliminary set of questions, we task GPT-4 with generating SQL queries that correspond to these questions. The GPT-4 model is instructed to construct modular SQL queries that are easy to read and utilize common table expressions (CTEs).\\
Subsequently, the generated SQL queries are executed on both data warehouses. In case of any errors or absence of results, a self-healing loop is employed to rectify the issue, with a maximum of five retry attempts. If the issue remains unresolved after these attempts, the corresponding question is removed from the dataset.\\
Next, GPT-4 is instructed to filter the resolved questions, retaining only those for which the dimensions of the dataframe, the column names, and the initial five rows of the results logically align with the corresponding question.\\
Additionally, GPT-4 is requested to reformulate each retained question up to four times, resulting in five distinct questions that can each be answered using the same SQL query. This approach ensures a broader diversity in the syntax and semantics of the questions, thereby approximating real-world usage.\\
The resultant dataset comprises 822 unique pairs of question-SQL query. After applying data augmentation through question rewrites, the total dataset encompasses 3732 such pairs (Figure \ref{flow}).
The dataset is subsequently partitioned into training (80\%) and testing sets.



\subsection{Models and finetuning}
 Our base model selection comprises three models of varying parameter sizes, namely Starcoder Plus \cite{li2023starcoder}, Code-Llama (the 13 billion parameters version) \cite{rozière2023code}, and Mistral-7b \cite{jiang2023mistral}.\\
Starcoder Plus is a 15.5 billion parameter model. It boasts an extensive 8K context length and is proficient in infilling tasks. It efficiently executes large-batch inferences, thanks to its multi-query attention mechanism. This model is a product of the BigCode community – a collaborative, open-science initiative dedicated to the ethical advancement of Large Language Models specifically designed for code generation and manipulation. It is based on StarCoderBase, which was trained on 1 trillion tokens sourced from The Stack \cite{kocetkov2022stack}, a large collection of permissively licensed GitHub repositories with inspection tools and an opt-out process. The model then underwent further fine-tuning specifically for the English language.\\
Code LLama is a series of large language models specifically designed for code, based on Llama 2\cite{touvron2023llama}. These models exhibit competitive performance in comparison to other open models, offering features such as infilling, handling extensive input contexts, and demonstrating zero-shot instruction following capabilities for various programming tasks.\\
Mistral is a language model consisting of 7 billion parameters, utilizing grouped-query attention (GQA) for efficient inference and sliding window attention (SWA) to manage sequences of varying lengths while minimizing inference costs. In comparison to the 13-billion-parameter Llama 2 model, Mistral demonstrates superior performance across all assessed benchmarks. \\
 The finetuning set-up is the same for each models, with only small changes in parameters. \\
Due to resource constraints, we opted to employ the LoRa (Low Rank Adaptors) technique for all three models\cite{Hu2021LoRALA}. This approach enabled us to conduct each fine-tuning on a single A100 Nvidia GPU with a memory capacity of 40GB.\\
Moreover, during the training process, we disregarded the prompt loss and focused solely on the completion tokens. By doing so, the model was able to concentrate its efforts on generating the correct SQL query completion rather than the provided prompt text, which served merely as a context.\\
Furthermore, we refrained from packing the prompt-completion pairs into fixed token length sequences. Instead, we kept them separate, preventing any potential loss of information and ensuring that the prompt contained the full context.


\section{Results}
 Assessing the validity of SQL queries can be a challenging task. Employing a model such as GPT-4 as a judge presents several issues related to accuracy. Moreover, a single question may be answered correctly in multiple ways, and the number and aliases of columns in the results can vary. To address these concerns, we have devised a deterministic evaluation method that takes these factors into account.
 
\subsection{Evaluation methodology}
 In order to assess the results, we elected to concentrate on the data produced by the SQL query rather than the query itself. To this end, we executed the ground truth query and saved the output. Subsequently, we compared the results to those returned by the generated query. Our initial step involved verifying that both queries yielded the same number of rows. Subsequently, we confirmed that the ground truth columns were present in the generated query's dataframe, without considering column names. This comparison was conducted in a greedy manner. Lastly, we accounted for the possibility that the rows might be returned in a different order with sorting.
\subsection{Results}
\begin{figure}
  \begin{subfigure}{0.5\textwidth}
    \centering
    \adjustbox{valign=m,vspace=1cm}{\includegraphics[width=0.9\linewidth]{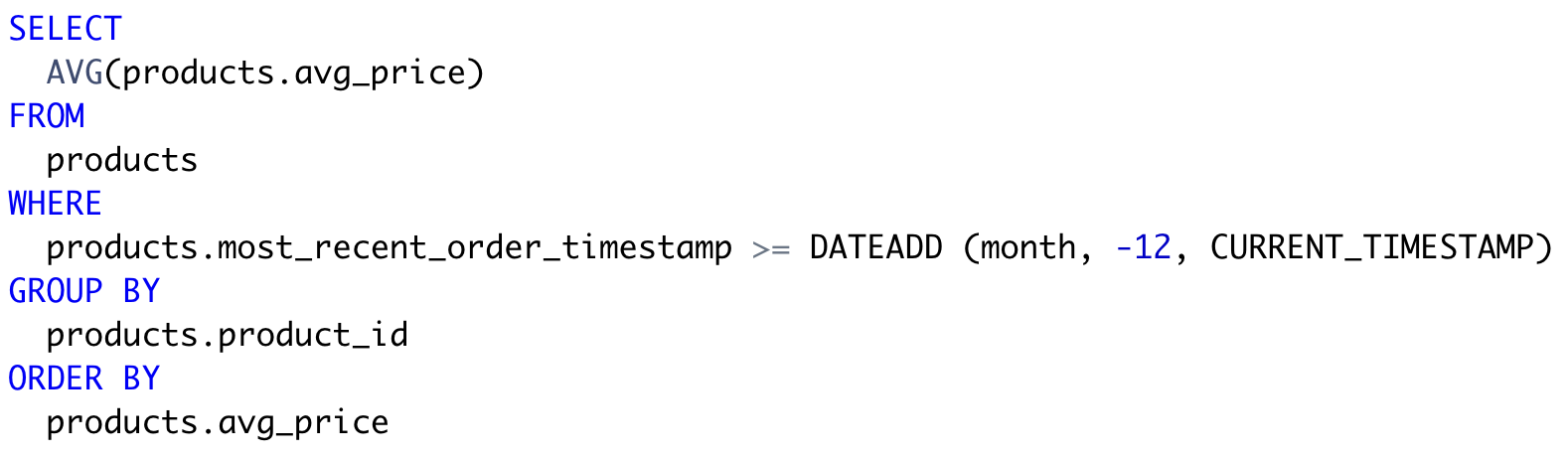}}
    \caption{Generated SQL query before finetuning}
  \end{subfigure}%
  \begin{subfigure}{0.5\textwidth}
    \centering
    \adjustbox{valign=m}{\includegraphics[width=0.9\linewidth]{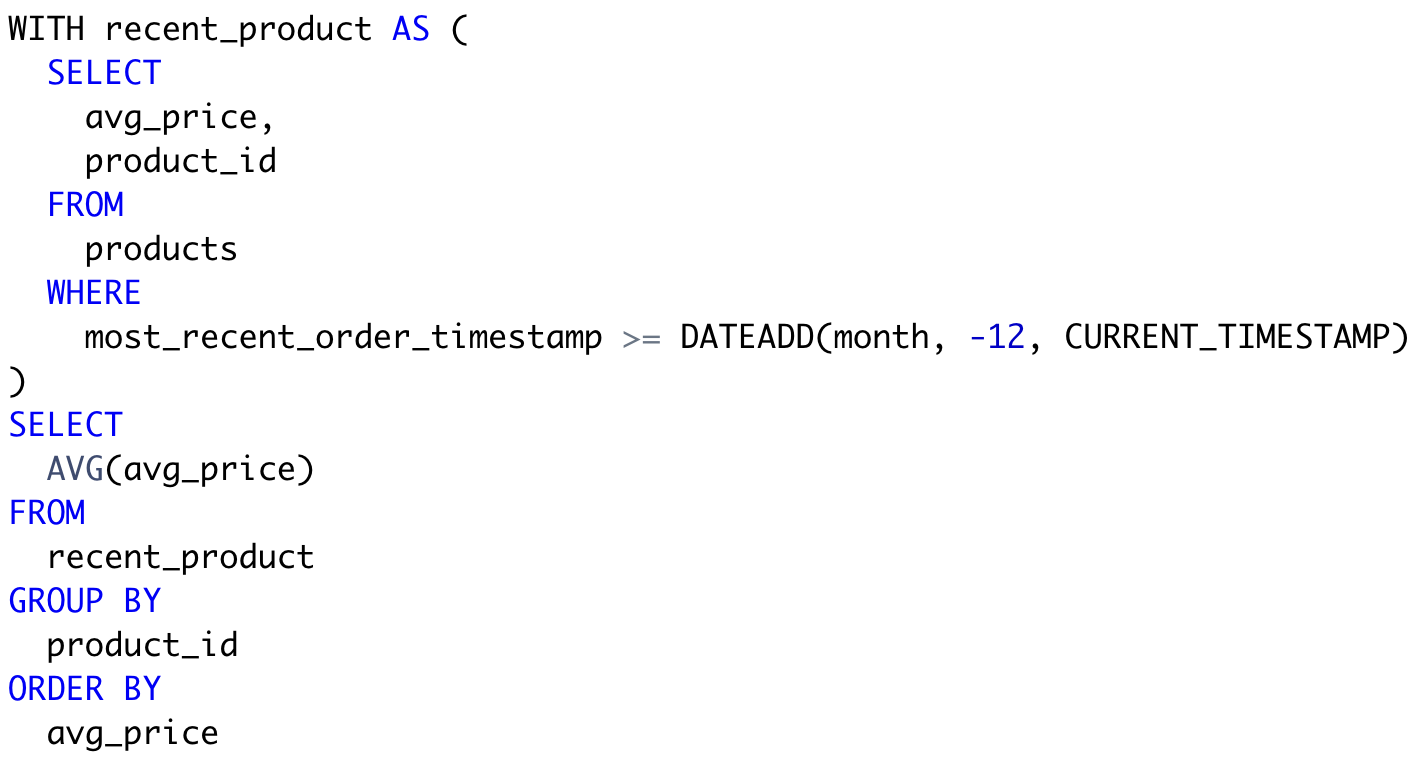}}
    \caption{Generated SQL query after finetuning}
  \end{subfigure}
  \caption{Comparison of SQL syntax before and after finetuning on a simple example}
  \label{double}
\end{figure}

 Table \ref{res} presents a comparison of the fine-tuned models' results in a zero-shot setting for each SQL dialect, relative to the zero-shot performance of GPT-4, as applied to the generated retail data. Furthermore, the models are executed using quantization with int8 precision. We provide the average duration of result generation, the success rate which is the percentage of SQL queries that return any results and the accuracy, meaning the percentage of SQL queries that return correct results out of the whole test dataset. 
 We also noticed that the finetuned models are much better at writing modular SQL queries that are easy to read and utilize common table expressions (CTEs) (Figure \ref{double})
 \begin{table}
     \centering
     \begin{tabular}{|c|c|c|c|}
     \hline
       Models   & Query Duration (Avg.)  & Success Rate (\%) & Accuracy Rate (\%)\\
    \hline
          GPT-4 (Snowflake)&  47.44s&87.36\%  &45.64\% \\
          GPT-4 (Google SQL)&47.40s  &90.97\%  &48.89\% \\
          Starcoder (Snowflake)&  16.6s&92.93\%  &62.43\% \\
          Starcoder (Google SQL)&19.32s  &87.94\%  &51.05\% \\
          Code-LLama (Snowflake)&  5.47s&98.73\%  &81.58\% \\
          Code-LLama (Google SQL)&6.46s  &97.76\%  &82.66\% \\
          Mistral (Snowflake)&  5.67s&96.45\%  &79.60\% \\
          Code-LLama (Google SQL)&6.50s  &96.08\%  &79.44\% \\
          \hline
     \end{tabular}
     \caption{Benchmark results on the generated dataset, all models except GPT-4 are finetuned}
     \label{res}
 \end{table}

\subsection{Discussion}

The development of models specialized for tasks like SQL query generation clearly benefits response latency and computational efficiency. Our investigation revealed that while the Starcoder model exhibited slower response times compared to Mistral and Code-LLama, this was largely due to the absence of certain inference optimization techniques like EETQ quantization\cite{EETQ} at the time of testing. Nevertheless, the fine-tuned models demonstrated a significant performance edge over the zero-shot GPT-4 model, with Code-LLama achieving the highest accuracy rates for both the Snowflake and Google SQL dialects.

The ability of these models to produce syntactically correct SQL queries that are executable by the database system is a testament to the effectiveness of the fine-tuning process. However, the specialized nature of these models does raise questions about their generalizability. Although the models' high accuracy rates are promising, their performance may decline when faced with SQL queries outside the training dataset's scope. This potential overfitting to the training data is a critical consideration for real-world applications, where the diversity of queries can be vast.

Furthermore, the similarity between the training and test datasets may inflate the measured performance of the models. While the test data was deliberately chosen to represent realistic retail SQL tasks, the possibility that the models have learned to excel on a narrow set of problems cannot be ignored. This underscores the need for comprehensive testing on a variety of datasets to fully understand the models' capabilities and limitations.

Despite these concerns, the results of this study are encouraging, particularly for domain-specific applications where the ability to quickly and accurately generate SQL queries can significantly impact the accessibility and usability of database systems. The fine-tuned models' proficiency in understanding and applying complex SQL syntax and structures suggests that with further refinement and broader testing, we can enhance their applicability and reliability for a wider array of SQL-related tasks.


\section{Conclusion}

This paper contributes to the field of natural language processing and database querying by detailing the process of fine-tuning open-source large language models (LLMs) to accurately convert natural language questions into SQL queries in a retail context. Our approach, which leverages a synthetic dataset and the LoRa technique, has been shown to significantly enhance the performance of models like Starcoder Plus, Code-Llama, and Mistral, particularly when compared to the zero-shot performance of GPT-4. Code-Llama, in particular, demonstrated exceptional accuracy, surpassing 80\% in both Snowflake SQL and GoogleSQL dialects.\\
Through the course of our discussion, we have acknowledged the potential limitations of our study, including the possibility of overfitting and the lack of generalization beyond the specific dataset and SQL dialects used. These issues highlight the importance of careful dataset curation and rigorous model evaluation in future research.\\
Moving forward, there is a clear opportunity to expand this research to encompass a wider range of domains and SQL dialects, and to explore the integration of these models into end-user applications. Striking a balance between accuracy through specialization and broader applicability through generalization remains a key challenge in deploying these models.\\
By focusing on the intersection of machine learning, AI, and user accessibility, this study not only progresses the technical capabilities of Text-to-SQL models but also emphasizes the practical benefits of making complex data systems more accessible to non-experts. Continued refinement of these models brings us closer to a future with lower barriers to data insights, enabling a wider audience to harness the power of data analytics.\\
\bibliographystyle{unsrt}  
\bibliography{references}

\begin{thebibliography}{10}

\bibitem{cai2017}
Ruichu Cai, Boyan Xu, Zhenjie Zhang, Xiaoyan Yang, Zijian Li, and Zhihao Liang.
\newblock An encoder-decoder framework translating natural language to database queries.
\newblock In {\em International Joint Conference on Artificial Intelligence}, 2017.

\bibitem{Qin2022ASO}
Bowen Qin, Binyuan Hui, Lihan Wang, Min Yang, Jinyang Li, Binhua Li, Ruiying Geng, Rongyu Cao, Jian Sun, Luo Si, Fei Huang, and Yongbin Li.
\newblock A survey on text-to-sql parsing: Concepts, methods, and future directions.
\newblock {\em ArXiv}, abs/2208.13629, 2022.

\bibitem{Xu2017SQLNetGS}
Xiaojun Xu, Chang Liu, and Dawn~Xiaodong Song.
\newblock Sqlnet: Generating structured queries from natural language without reinforcement learning.
\newblock {\em ArXiv}, abs/1711.04436, 2017.

\bibitem{Yaghmazadeh2017SQLizerQS}
Navid Yaghmazadeh, Yuepeng Wang, Işıl Dillig, and Thomas Dillig.
\newblock Sqlizer: query synthesis from natural language.
\newblock {\em Proceedings of the ACM on Programming Languages}, 1:1 -- 26, 2017.

\bibitem{Yu2018TypeSQLKT}
Tao Yu, Zifan Li, Zilin Zhang, Rui Zhang, and Dragomir~R. Radev.
\newblock Typesql: Knowledge-based type-aware neural text-to-sql generation.
\newblock In {\em North American Chapter of the Association for Computational Linguistics}, 2018.

\bibitem{Yu2018SpiderAL}
Tao Yu, Rui Zhang, Kai-Chou Yang, Michihiro Yasunaga, Dongxu Wang, Zifan Li, James Ma, Irene~Z Li, Qingning Yao, Shanelle Roman, Zilin Zhang, and Dragomir~R. Radev.
\newblock Spider: A large-scale human-labeled dataset for complex and cross-domain semantic parsing and text-to-sql task.
\newblock {\em ArXiv}, abs/1809.08887, 2018.

\bibitem{Zhong2017Seq2SQLGS}
Victor Zhong, Caiming Xiong, and Richard Socher.
\newblock Seq2sql: Generating structured queries from natural language using reinforcement learning.
\newblock {\em ArXiv}, abs/1709.00103, 2017.

\bibitem{Zhong2020GroundedAF}
Victor Zhong, Mike Lewis, Sida~I. Wang, and Luke Zettlemoyer.
\newblock Grounded adaptation for zero-shot executable semantic parsing.
\newblock In {\em Conference on Empirical Methods in Natural Language Processing}, 2020.

\bibitem{Gao2023TexttoSQLEB}
Dawei Gao, Haibin Wang, Yaliang Li, Xiuyu Sun, Yichen Qian, Bolin Ding, and Jingren Zhou.
\newblock Text-to-sql empowered by large language models: A benchmark evaluation.
\newblock {\em ArXiv}, abs/2308.15363, 2023.

\bibitem{Raffel2019ExploringTL}
Colin Raffel, Noam~M. Shazeer, Adam Roberts, Katherine Lee, Sharan Narang, Michael Matena, Yanqi Zhou, Wei Li, and Peter~J. Liu.
\newblock Exploring the limits of transfer learning with a unified text-to-text transformer.
\newblock {\em J. Mach. Learn. Res.}, 21:140:1--140:67, 2019.

\bibitem{Scholak2021PICARDPI}
Torsten Scholak, Nathan Schucher, and Dzmitry Bahdanau.
\newblock Picard: Parsing incrementally for constrained auto-regressive decoding from language models.
\newblock {\em ArXiv}, abs/2109.05093, 2021.

\bibitem{OpenAI2023GPT4TR}
OpenAI.
\newblock Gpt-4 technical report.
\newblock {\em ArXiv}, abs/2303.08774, 2023.

\bibitem{Anil2023PaLM2T}
Rohan Anil, Andrew~M. Dai, Orhan Firat, Melvin Johnson, Dmitry Lepikhin, Alexandre~Tachard Passos, Siamak Shakeri, Emanuel Taropa, Paige Bailey, Z.~Chen, Eric Chu, J.~Clark, Laurent~El Shafey, Yanping Huang, Kathleen~S. Meier-Hellstern, Gaurav Mishra, Erica Moreira, Mark Omernick, Kevin Robinson, Sebastian Ruder, Yi~Tay, Kefan Xiao, Yuanzhong Xu, Yujing Zhang, Gustavo~Hernandez Abrego, Junwhan Ahn, Jacob Austin, Paul Barham, Jan~A. Botha, James Bradbury, Siddhartha Brahma, Kevin~Michael Brooks, Michele Catasta, Yongzhou Cheng, Colin Cherry, Christopher~A. Choquette-Choo, Aakanksha Chowdhery, C~Cr{\'e}py, Shachi Dave, Mostafa Dehghani, Sunipa Dev, Jacob Devlin, M.~C. D'iaz, Nan Du, Ethan Dyer, Vladimir Feinberg, Fan Feng, Vlad Fienber, Markus Freitag, Xavier Garc{\'i}a, Sebastian Gehrmann, Lucas Gonz{\'a}lez, Guy Gur-Ari, Steven Hand, Hadi Hashemi, Le~Hou, Joshua Howland, An~Ren Hu, Jeffrey Hui, Jeremy Hurwitz, Michael Isard, Abe Ittycheriah, Matthew Jagielski, Wen~Hao Jia, Kathleen Kenealy, Maxim Krikun,
  Sneha Kudugunta, Chang Lan, Katherine Lee, Benjamin Lee, Eric Li, Mu-Li Li, Wei Li, Yaguang Li, Jun~Yu Li, Hyeontaek Lim, Han Lin, Zhong-Zhong Liu, Frederick Liu, Marcello Maggioni, Aroma Mahendru, Joshua Maynez, Vedant Misra, Maysam Moussalem, Zachary Nado, John Nham, Eric Ni, Andrew Nystrom, Alicia Parrish, Marie Pellat, Martin Polacek, Alex Polozov, Reiner Pope, Siyuan Qiao, Emily Reif, Bryan Richter, Parker Riley, Alexandra Ros, Aurko Roy, Brennan Saeta, Rajkumar Samuel, Renee~Marie Shelby, Ambrose Slone, Daniel Smilkov, David~R. So, Daniela Sohn, Simon Tokumine, Dasha Valter, Vijay Vasudevan, Kiran Vodrahalli, Xuezhi Wang, Pidong Wang, Zirui Wang, Tao Wang, John Wieting, Yuhuai Wu, Ke~Xu, Yunhan Xu, Lin~Wu Xue, Pengcheng Yin, Jiahui Yu, Qiaoling Zhang, Steven Zheng, Ce~Zheng, Wei Zhou, Denny Zhou, Slav Petrov, and Yonghui Wu.
\newblock Palm 2 technical report.
\newblock {\em ArXiv}, abs/2305.10403, 2023.

\bibitem{Wei2022EmergentAO}
Jason Wei, Yi~Tay, Rishi Bommasani, Colin Raffel, Barret Zoph, Sebastian Borgeaud, Dani Yogatama, Maarten Bosma, Denny Zhou, Donald Metzler, Ed~Huai hsin Chi, Tatsunori Hashimoto, Oriol Vinyals, Percy Liang, Jeff Dean, and William Fedus.
\newblock Emergent abilities of large language models.
\newblock {\em Trans. Mach. Learn. Res.}, 2022, 2022.

\bibitem{Chen2021EvaluatingLL}
Mark Chen, Jerry Tworek, Heewoo Jun, Qiming Yuan, Henrique Ponde, Jared Kaplan, Harrison Edwards, Yura Burda, Nicholas Joseph, Greg Brockman, Alex Ray, Raul Puri, Gretchen Krueger, Michael Petrov, Heidy Khlaaf, Girish Sastry, Pamela Mishkin, Brooke Chan, Scott Gray, Nick Ryder, Mikhail Pavlov, Alethea Power, Lukasz Kaiser, Mohammad Bavarian, Clemens Winter, Philippe Tillet, Felipe~Petroski Such, David~W. Cummings, Matthias Plappert, Fotios Chantzis, Elizabeth Barnes, Ariel Herbert-Voss, William~H. Guss, Alex Nichol, Igor Babuschkin, S.~Arun Balaji, Shantanu Jain, Andrew Carr, Jan Leike, Joshua Achiam, Vedant Misra, Evan Morikawa, Alec Radford, Matthew~M. Knight, Miles Brundage, Mira Murati, Katie Mayer, Peter Welinder, Bob McGrew, Dario Amodei, Sam McCandlish, Ilya Sutskever, and Wojciech Zaremba.
\newblock Evaluating large language models trained on code.
\newblock {\em ArXiv}, abs/2107.03374, 2021.

\bibitem{Liu2023ACE}
Aiwei Liu, Xuming Hu, Lijie Wen, and Philip~S. Yu.
\newblock A comprehensive evaluation of chatgpt's zero-shot text-to-sql capability.
\newblock {\em ArXiv}, abs/2303.13547, 2023.

\bibitem{Sun2023SQLPaLMIL}
Ruoxi Sun, Sercan~{\"O}. Arik, Hootan Nakhost, Hanjun Dai, Rajarishi~S. Sinha, Pengcheng Yin, and Tomas Pfister.
\newblock Sql-palm: Improved large language model adaptation for text-to-sql.
\newblock {\em ArXiv}, abs/2306.00739, 2023.

\bibitem{defog}
Wendy~Aw Wong Jing~Ping and Rishabh Srivastava.
\newblock Open-sourcing sqlcoder2-15b and sqlcoder-7b, 2023.
\newblock https://defog.ai/blog/open-sourcing-sqlcoder2-7b/.

\bibitem{numbersstation2023NSText2SQL}
Numbers~Station Labs.
\newblock Nstext2sql: An open source text-to-sql dataset for foundation model training, July 2023.
\newblock https://github.com/NumbersStationAI/NSQL.

\bibitem{Li2023CanLA}
Jinyang Li, Binyuan Hui, Ge~Qu, Binhua Li, Jiaxi Yang, Bowen Li, Bailin Wang, Bowen Qin, Rongyu Cao, Ruiying Geng, Nan Huo, Chenhao Ma, Kevin~C. Chang, Fei Huang, Reynold Cheng, and Yongbin Li.
\newblock Can llm already serve as a database interface? a big bench for large-scale database grounded text-to-sqls.
\newblock {\em ArXiv}, abs/2305.03111, 2023.

\bibitem{sqlglot}
Toby Mao.
\newblock Sqlglot, 2023.
\newblock https://sqlglot.com/sqlglot.html.

\bibitem{li2023starcoder}
Raymond Li, Loubna~Ben Allal, Yangtian Zi, Niklas Muennighoff, Denis Kocetkov, Chenghao Mou, Marc Marone, Christopher Akiki, Jia Li, Jenny Chim, Qian Liu, Evgenii Zheltonozhskii, Terry~Yue Zhuo, Thomas Wang, Olivier Dehaene, Mishig Davaadorj, Joel Lamy-Poirier, João Monteiro, Oleh Shliazhko, Nicolas Gontier, Nicholas Meade, Armel Zebaze, Ming-Ho Yee, Logesh~Kumar Umapathi, Jian Zhu, Benjamin Lipkin, Muhtasham Oblokulov, Zhiruo Wang, Rudra Murthy, Jason Stillerman, Siva~Sankalp Patel, Dmitry Abulkhanov, Marco Zocca, Manan Dey, Zhihan Zhang, Nour Fahmy, Urvashi Bhattacharyya, Wenhao Yu, Swayam Singh, Sasha Luccioni, Paulo Villegas, Maxim Kunakov, Fedor Zhdanov, Manuel Romero, Tony Lee, Nadav Timor, Jennifer Ding, Claire Schlesinger, Hailey Schoelkopf, Jan Ebert, Tri Dao, Mayank Mishra, Alex Gu, Jennifer Robinson, Carolyn~Jane Anderson, Brendan Dolan-Gavitt, Danish Contractor, Siva Reddy, Daniel Fried, Dzmitry Bahdanau, Yacine Jernite, Carlos~Muñoz Ferrandis, Sean Hughes, Thomas Wolf, Arjun Guha, Leandro von
  Werra, and Harm de~Vries.
\newblock Starcoder: may the source be with you!, 2023.

\bibitem{rozière2023code}
Baptiste Rozière, Jonas Gehring, Fabian Gloeckle, Sten Sootla, Itai Gat, Xiaoqing~Ellen Tan, Yossi Adi, Jingyu Liu, Tal Remez, Jérémy Rapin, Artyom Kozhevnikov, Ivan Evtimov, Joanna Bitton, Manish Bhatt, Cristian~Canton Ferrer, Aaron Grattafiori, Wenhan Xiong, Alexandre Défossez, Jade Copet, Faisal Azhar, Hugo Touvron, Louis Martin, Nicolas Usunier, Thomas Scialom, and Gabriel Synnaeve.
\newblock Code llama: Open foundation models for code, 2023.

\bibitem{jiang2023mistral}
Albert~Q. Jiang, Alexandre Sablayrolles, Arthur Mensch, Chris Bamford, Devendra~Singh Chaplot, Diego de~las Casas, Florian Bressand, Gianna Lengyel, Guillaume Lample, Lucile Saulnier, Lélio~Renard Lavaud, Marie-Anne Lachaux, Pierre Stock, Teven~Le Scao, Thibaut Lavril, Thomas Wang, Timothée Lacroix, and William~El Sayed.
\newblock Mistral 7b, 2023.

\bibitem{kocetkov2022stack}
Denis Kocetkov, Raymond Li, Loubna~Ben Allal, Jia Li, Chenghao Mou, Carlos~Muñoz Ferrandis, Yacine Jernite, Margaret Mitchell, Sean Hughes, Thomas Wolf, Dzmitry Bahdanau, Leandro von Werra, and Harm de~Vries.
\newblock The stack: 3 tb of permissively licensed source code, 2022.

\bibitem{touvron2023llama}
Hugo Touvron, Louis Martin, Kevin Stone, Peter Albert, Amjad Almahairi, Yasmine Babaei, Nikolay Bashlykov, Soumya Batra, Prajjwal Bhargava, Shruti Bhosale, Dan Bikel, Lukas Blecher, Cristian~Canton Ferrer, Moya Chen, Guillem Cucurull, David Esiobu, Jude Fernandes, Jeremy Fu, Wenyin Fu, Brian Fuller, Cynthia Gao, Vedanuj Goswami, Naman Goyal, Anthony Hartshorn, Saghar Hosseini, Rui Hou, Hakan Inan, Marcin Kardas, Viktor Kerkez, Madian Khabsa, Isabel Kloumann, Artem Korenev, Punit~Singh Koura, Marie-Anne Lachaux, Thibaut Lavril, Jenya Lee, Diana Liskovich, Yinghai Lu, Yuning Mao, Xavier Martinet, Todor Mihaylov, Pushkar Mishra, Igor Molybog, Yixin Nie, Andrew Poulton, Jeremy Reizenstein, Rashi Rungta, Kalyan Saladi, Alan Schelten, Ruan Silva, Eric~Michael Smith, Ranjan Subramanian, Xiaoqing~Ellen Tan, Binh Tang, Ross Taylor, Adina Williams, Jian~Xiang Kuan, Puxin Xu, Zheng Yan, Iliyan Zarov, Yuchen Zhang, Angela Fan, Melanie Kambadur, Sharan Narang, Aurelien Rodriguez, Robert Stojnic, Sergey Edunov, and Thomas
  Scialom.
\newblock Llama 2: Open foundation and fine-tuned chat models, 2023.

\bibitem{Hu2021LoRALA}
J.~Edward Hu, Yelong Shen, Phillip Wallis, Zeyuan Allen-Zhu, Yuanzhi Li, Shean Wang, and Weizhu Chen.
\newblock Lora: Low-rank adaptation of large language models.
\newblock {\em ArXiv}, abs/2106.09685, 2021.

\bibitem{EETQ}
Eetq, 2023.
\newblock https://github.com/NetEase-FuXi/EETQ.

\end{thebibliography}

\end{document}